# Controllability Gramian Spectra of Random Networks

Victor M. Preciado* & M. Amin Rahimian

*Abstract*—We propose a theoretical framework to study the eigenvalue spectra of the controllability Gramian of systems with random state matrices, such as networked systems with a random graph structure. Using random matrix theory, we provide expressions for the moments of the eigenvalue distribution of the controllability Gramian. These moments can then be used to derive useful properties of the eigenvalue distribution of the Gramian (in some cases, even closed-form expressions for the distribution). We illustrate this framework by considering system matrices derived from common random graph and matrix ensembles, such as the Wigner ensemble, the Gaussian Orthogonal Ensemble (GOE), and random regular graphs. Subsequently, we illustrate how the eigenvalue distribution of the Gramian can be used to draw conclusions about the energy required to control random system.

## I. INTRODUCTION

Controllability is a classical tool in systems theory, as developed by Kalman et al. [1]. More recently, controllability of complex networked systems has attracted much attention, beginning with the study of leader-follower multi-agent systems in [2], [3], [4], [5], [6], followed by several works considering the network controllability problem in special topologies such as paths [7], circulant networks [8], grids [9], distance regular graphs [10], and networks-of-networks [11], as well as works which employ the concept of structural controllability in networked settings such as [12], [13], [14].

The concept of controllability is closely related to the minimum energy needed to steer a linear system through its state space [1], [15], [16]. In particular, the minimum input energy needed to drive the system from an initial to a desired state can be expressed in terms of the inverse of the controllability Gramian matrix, which we denote by $W$ [17], [16]. This relationship has motivated several Gramian-based metrics to quantify the energy required to steer a system, such as the trace of the Gramian inverse $\text{tr}(W^{-1})$, the inverse of the minimum eigenvalue $1/\lambda_{\min}(W)$, or $-\log(\det(W))$, [16]. These measures have found recent applications in the analysis and design of networked control systems [18], [19], [20]. These energy control metrics are closely related to the Gramian eigenvalue spectrum, which is the central topic of our work.

In this paper, we focus our attention on the eigenvalue spectrum of the controllability Gramian of systems with random state matrices. In this case, the controllability Gramian is also a random (symmetric) matrix with random (real) eigenvalues following a probability distribution, called the *Gramian spectral distribution*. The study of the eigenvalues of random matrices started in the pioneering work of Wigner [21], [22], Mehta and Gaudin [23], and Dyson [24], and has attracted a great deal of attention since then. Over the last decades, random matrix theory has become a mature field and found applications in a wide variety of disciplines [25], [26]. A particularly relevant application of random matrix theory can be found in the study of the eigenvalues of random graph models [27], [28], [29], [30]. In this direction, many graph-related matrices—such as the adjacency, the Laplacian, and the normalized Laplacian—have been widely studied in the literature. In contrast, theoretical results about the eigenvalues of the controllability/observability Gramians of random graphs and matrices are very scarce.

The are only a few works that attempt to address some aspects of this problem [31], [32], [33]. The author in [31] introduces a stochastic notion of controllability for jump systems, where the state and input matrices vary between $N$ designated matrix pairs and the variations follow a Markov process. Accordingly, the linear jump system is controllable if transitions between any two states occur with positive probability and in an almost surely finite length of time; variations of this definition when the transitions occur with probability one and the required time-span has finite expectation are also considered, and an algebraic criterion for stochastic controllability is expressed as a rank condition on the $N$ state and input matrix pairs. Parallel results are derived in [32], but for the case where the state matrix is given by the Laplacian of a random graph process. In [33] the authors address related observability problems by investigating the recovery of sparse initial states using independent and randomly populated measurement matrices and under certain conditions on the state transition matrix.

In this paper, we study the Gramian spectral distribution of random systems, paying special attention to the Gramian *spectral moments* (defined as the moments of the eigenvalue distribution). In particular, we establish a connection between the spectral moments of the (random) state matrix of a system and the spectral moments of its (random) controllability Gramian. We then exploit this connection to derive closed-form expressions for the Gramian spectral distribution of important random matrices, such as Wigner matrices, the Gaussian Orthogonal Ensemble (GOE), or random regular graphs. From the Gramian spectral distribution, we then draw conclusions about the energy required to steer a random system and study how this energy is affected by the system parameters and the dimension of the state space.

The paper is organized as follows. Notation and preliminary results are provided in Section II. Our main results are stated in Section III and proofs are included in Appendix A. Illustrative examples and additional discussions are presented in Section IV and the paper is concluded in Section V.

* The authors are with the Department of Electrical and Systems Engineering, University of Pennsylvania, Philadelphia, PA 19104-6228 USA. (email: preciado@seas.upenn.edu). This work was supported by the National Science Foundation under grants CNS-1302222 and IIS-1447470.

## II. NOTATION & PRELIMINARY RESULTS

*Notation.:* Throughout the paper, $\mathbb{R}$ is the set of real numbers, boldface letters denote random variables, $\mathbb{E}\{\cdot\}$ is the expectation operator, vectors are denoted by a bar over their respective lower-case letters, matrices are denoted by upper case letters, $^T$ denotes matrix transpose and $I_n$ is the $n \times n$ identity matrix. We use $\sim$ to indicate asymptotic equivalence as $n \to \infty$, i.e., $f(n) \sim g(n)$ indicates that $\lim_{n\to\infty} \frac{f(n)}{g(n)} = 1$.

### A. The Method of Moments

Let us consider a random system characterized by the matrix pair $(\mathbf{A}_n, B_n)$, where $\mathbf{A}_n$ is a random $n \times n$ state matrix and $B_n$ is an input matrix. We investigate the limiting distribution of the eigenvalues of the controllability Gramian associated with the matrix pair $(\mathbf{A}_n, B_n)$ as $n \to \infty$. It is noted in [34], [35] that the required control energy can increase unbounded with the growing network size, unless $B_n = I_n$; hence, we restrict attention to the case $B_n = I_n$. We also assume that $\mathbf{A}_n$ is symmetric[1]. Given an $n \times n$ symmetric random matrix $\mathbf{C}$, let $\lambda_1(C) \leq \lambda_2(C) \leq \ldots \leq \lambda_n(C)$ be $n$ random variables representing the $n$ real eigenvalues of $\mathbf{A}_n$. We consider the random probability measure $\mathcal{L}_\mathbf{C}\{\cdot\} = \frac{1}{n}\sum_{i=1}^n \delta_{\lambda_i(\mathbf{C})}\{\cdot\}$, $\delta_{\lambda_i(\mathbf{C})}$ being Dirac's delta function centered at $\lambda_i(\mathbf{C})$, as the random probability measure on the real line that assigns mass uniformly to each of the $n$ eigenvalues of the random matrix $\mathbf{C}$. The corresponding distribution $\mathbf{F}_n(x) = \mathcal{L}_\mathbf{C}\{(-\infty, x]\} = \frac{1}{n}|\{i \in [n] : \lambda_i(\mathbf{C}) \leq x\}|$, is a random variable for each $x \in \mathbb{R}$ and is referred to as the *empirical spectral distribution* (ESD) for the random matrix $\mathbf{C}$. The $k$-th *spectral moments* of a random matrix $\mathbf{C}$ is the $k$-th moment of its spectral density. These moments can be written as $m_k(\mathbf{C}) := \frac{1}{n}\sum_{i=1}^n \lambda_i(\mathbf{C})^k$ [25]. In our derivations, we also make use of the *centralized spectral moments*, which are defined as $\widetilde{m}_k(\mathbf{C}) := \frac{1}{n}\sum_{i=1}^n (\lambda_i(\mathbf{C}) - m_1(\mathbf{C}))^k$.

In this paper, we will use the method of moments to derive the limiting spectral distributions of the controllability Gramians herein considered. According to this method, when dealing with compactly supported distributions (as will be the case for us), one can show that a sequence of empirical spectral distributions $\mathbf{F}_n(\cdot)$ converges in probability to some limiting distribution $F(\cdot)$ by showing that the sequence of expected moments for $\mathbf{F}_n(\cdot)$, given by $\{\int_0^{+\infty} x^k d\mathbf{F}_n(x), k \in \mathbb{N}\}$, converges point-wise (for every $k$) to the corresponding moments of $F(\cdot)$, given by $\{\int_0^{+\infty} x^k dF(x), k \in \mathbb{N}\}$ [26, Theorem 2.2.9 and Section 2.4.2].

In the following subsections, we describe two particular random matrices that will be useful to illustrate our results.

### B. Random Wigner Matrices: Semicircle Law

A Wigner matrix $\mathbf{A}_n$ is a random matrix where the entries above the main diagonal are i.i.d. zero-mean with identical variance $\sigma^2$. In his seminal papers [21], [22], Eugene Wigner proved that the empirical spectral distribution of $(1/\sqrt{n})\mathbf{A}_n$ converges (in probability) to the semicircular density given by

$$f_{SC}(x) = \frac{1}{2\pi\sigma^2}\sqrt{4\sigma^2 - x^2}, \quad (1)$$

for $x \in [-2\sigma, 2\sigma]$ and $f_{SC}(x) = 0$, otherwise. The even-order moment of the semi-circular distribution are given by $m_{2j}^{SC} = \frac{\sigma^{2j}}{j+1}\binom{2j}{j}$; odd-order moments are zero. In fact, for this result to hold we do not need the variables to be identically distributed and the result continues to hold for independent zero-mean entires with common variance $\sigma^2$ [36], under certain mild conditions on higher-order moments. An important special case of Wigner's random matrix is the Gaussian Orthogonal Ensemble (GOE), in which the off-diagonal entries are real Gaussian variables with mean zero and variance $\sigma^2$. The diagonal entries are real Gaussian variables with mean zero and variance $\sigma^2/2$. We will use this particular ensemble to illustrate our results in Section IV.

### C. Random Regular Graphs: McKay Law

Consider a random graph drawn uniformly over the space of all undirected $d$-regular graphs on $n$ vertices, where we assume $d > 2$. As $n \to \infty$, the empirical spectral distribution of the adjacency matrix of a random $d$-regular graph converges in probability to the McKay law [37], whose density is given by

$$f_{MK}(x) = \frac{d\sqrt{4(d-1) - x^2}}{2\pi(d^2 - x^2)},$$

for $|x| \leq 2\sqrt{d-1}$, and $f_{MK}(x) = 0$, otherwise. The moments of the McKay distribution are given by

$$m_k^{MK} = \sum_{r=1}^{k/2} \binom{k}{r}\frac{k - 2r + 1}{k - r + 1}(d-1)^r, \quad (2)$$

for $k$ even, and $0$ for $k$ odd. In our derivations, we shall also make use of the centralized spectral moments of the Laplacian matrix. Let $\mathbf{C}_n$ be the (random) adjacency matrix of the graph (i.e., $[\mathbf{C}_n]_{ij} = 1$ if nodes $i$ and $j$ are connected, and $0$ otherwise). For a random $d$-regular graph, the Laplacian matrix can be written as $\mathbf{L}_n = dI_n - \mathbf{C}_n$. One can easily prove that the centralized spectral moments of the Laplacian satisfy $\mathbb{E}\{\widetilde{m}_k(\mathbf{L}_n)\} \sim m_k^{MK}$ as $n \to \infty$, in the case of random $d$-regular graphs. This fact will be useful in future derivations.

## III. MOMENT-BASED CHARACTERIZATION OF THE GRAMIAN SPECTRUM

Consider a discrete-time (DT) linear time-invariant system $\overline{x}[t+1] = \mathbf{A}_n\overline{x}[t] + B_n\overline{u}[t]$, where $B_n = I_n$, $\mathbf{A}_n$ is an $n \times n$ random symmetric matrix, and $\overline{u}[t] \in \mathbb{R}^n$. Assuming $\mathbf{A}_n$ is Schur stable and symmetric, the discrete-time controllability Gramian is an $n \times n$ symmetric random matrix given by

$$\mathbf{W}_d = \sum_{\tau=0}^\infty \mathbf{A}_n^\tau (\mathbf{A}_n^T)^\tau = (I_n - \mathbf{A}_n^2)^{-1}.$$

---

[1] More general input and state matrices, such as random input matrices and non-symmetric state matrices, are considered in an upcoming extended version of this paper.

Similarly, for a continuous-time (CT) linear time-invariant system described by $\frac{d}{dt}\overline{x}(t) = \mathbf{A}_n\overline{x}(t) + \overline{u}(t)$, when $\mathbf{A}_n$ is an $n \times n$ Hurwitz stable, symmetric random matrix the continuous-time controllability Gramian $\mathbf{W}_c$ is given by

$$\mathbf{W}_c = \int_0^\infty e^{\mathbf{A}_n\tau} e^{\mathbf{A}_n^T \tau} d\tau = \int_0^\infty e^{2\mathbf{A}_n\tau} d\tau$$
$$= \mathbf{V}\text{diag}\left\{\int_0^\infty \exp\left(2\lambda_i(\mathbf{A}_n)\tau\right)d\tau\right\}_{i=1}^n \mathbf{V}^T, \quad (3)$$

where in last equality we use the fact that $\mathbf{A}_n$ is symmetric and, therefore, orthogonally diagonalizable.

In the following subsections, for both random discrete-time and continuous-time systems, we first provide theorems to relate the spectral moments of the state matrix $\mathbf{A}_n$ to the spectral moments of the controllability matrix. Since the spectral moments of $\mathbf{A}_n$ are well-understood for a number of random matrix ensembles, our result allows us to calculate the spectral moments of the controllability Gramian. Finally, using the method of moments, we will characterize and completely determine the limiting spectral distribution of the Gramian matrix for a number of well-known random matrix ensembles.

### A. Gramian of Random Discrete-Time Systems

In the following lemma, we provide a useful relationship between the moments of the state matrix $\mathbf{A}_n$ and the moments of the associated controllability gramian $\mathbf{W}_d$ for random discrete-time systems.

**Lemma 1** (Spectral Moments of DT Gramian). Let $\mathbf{A}_n$ be an $n \times n$ symmetric, Schur-stable state matrix with spectral moments given by $\mathbf{m}_k := m_k(\mathbf{A}_n)$ for $k \in \mathbb{N}$. Then, the spectral moments of the controllability Gramian of the discrete-time system $(\mathbf{A}_n, I_n)$ are given by

$$m_k(\mathbf{W}_d) = \sum_{j=0}^\infty \binom{j+k-1}{k-1}\mathbf{m}_{2j}. \quad (4)$$

We will prove the above lemma in Appendix A. Lemma 1 can be used to, for example, characterize the Gramian spectrum when the state matrix $\mathbf{A}_n$ is a Wigner random matrix. In particular, let $\mathbf{A}_n = (1/\sqrt{n})\mathbf{H}_n$ where $\mathbf{H}_n$ is a Wigner random matrix with the second moment satisfying $\sigma < 1/2$. Then, explicit expressions for the spectral moments of the Gramian and the Gramian spectrum are given in the following theorem:

**Theorem 1** (The Wigner Ensemble). Consider a DT system with state matrix $\mathbf{A}_n = \frac{1}{\sqrt{n}}\mathbf{H}_n$, where $\mathbf{H}_n$ is a Wigner random matrix with independent zero-mean entries and common variance $\sigma^2 < 1/4$, and the identity input matrix, $\mathbf{B}_n = I_n$. Then, the expected spectral moments of the controllability Gramian $\mathbf{W}_d$ are asymptotically given by

$$m_k(\mathbf{W}_d) \sim {}_2F_1(\frac{1}{2}, k; 2, 4\sigma^2), \quad (5)$$

where ${}_2F_1(a,b;c;z)$ is the hyper-geometric function[2]. Furthermore, the limiting spectral distribution of $\mathbf{W}_d$ is characterized by the following density function

$$f_W(x) = \begin{cases} \frac{1}{2\pi\sigma^2 x^2}\sqrt{\frac{4\sigma^2 x - x + 1}{x-1}}, & \text{if } x \in [1, \frac{1}{1-4\sigma^2}]. \\ 0, & \text{otherwise.} \end{cases} \quad (6)$$

While proving Theorem 1 in Appendix B, we provide a series of tools that can be directly applied to many other random matrix ensembles, as soon as they are Hurwitz stable. In the particular case treated in Theorem 1, the assumption $\sigma^2 < 1/4$ guarantees the matrix ensemble $\mathbf{A}_n$ to be Hurwitz stable. We now develop parallel results for random systems in the continuous-time case.

### B. Gramian of Random Continuous-Time Systems

Similarly to Subsection III-A, we first provide a lemma (proved in Appendix C) that allows us to calculate the asymptotic spectral moments of the Gramian from those of the state matrix $\mathbf{A}_n$. We then apply this result to study the Gramian of a system whose state matrix is related to the Laplacian of a random regular graph.

**Lemma 2** (Spectral Moments of the CT Gramian). Let $\mathbf{A}_n$ be an $n \times n$ symmetric, Schur-stable state matrix with spectral moments given by $\mathbf{m}_k := m_k(\mathbf{A}_n)$ and centralized spectral moments given by $\widetilde{\mathbf{m}}_k := \widetilde{m}_k(\mathbf{A}_n)$ for $k \in \mathbb{N}$. Then, the spectral moments of the continuous-time controllability Gramian $\mathbf{W}_c$ are given by:

$$m_k(\mathbf{W}_c) = \frac{1}{(-2)^k}\sum_{j=0}^\infty \binom{-k}{j}\frac{\widetilde{\mathbf{m}}_j}{(\mathbf{m}_1)^{k+j}}. \quad (7)$$

In what follows, we illustrate the usage of Lemma 2 by considering the Laplacian matrix of a random $d$-regular graph, described in Subsection II-C. Since the Laplacian matrix is marginally stable (has an eigenvalue at zero), the Gramian of the continuous-time system $(-\mathbf{L}_n, I_n)$ is ill-defined. To overcome this issue, we use a stabilized version of the Laplacian dynamics, where the state matrix is given by $\mathbf{A}_n = -\widetilde{\mathbf{L}}_n := -\mathbf{L}_n - \frac{d}{n}\mathbb{1}\mathbb{1}^T$. Notice that this matrix is not singular for connected graphs; thus, the Gramian of the system $(-\widetilde{\mathbf{L}}_n, I_n)$ is well-defined and satisfies (3). The following theorem, proved in Appendix D, provides explicit expressions for the spectral moments of the Gramian and the Gramian spectrum.

**Theorem 2** (Random Regular Graphs). Consider a CT system with input matrix $I_n$ and state matrix $\mathbf{A}_n = -\widetilde{\mathbf{L}}_n$, where $\widetilde{\mathbf{L}}_n$ is the stabilized Laplacian matrix of a random $d$-regular graph with $n$ nodes and $d \geq 3$. Then, the asymptotic expected spectral moments of the controllability Gramian $\mathbf{W}_c$ are given

---

[2] The hyper-geometric function is defined for $|z| < 1$ by the power series ${}_2F_1(q,b;c;z) = \sum_{j=0}^\infty \frac{(q)_j(b)_j}{(c)_j}\frac{z^j}{j!}$, where $(q)_j = q(q+1)\ldots(q+j-1)$.

by

$$m_k(\mathbf{W}_c) \sim \frac{1}{(2d)^k} \sum_{j=0}^{\infty} \binom{2j+k-1}{2j} d^{-2j} m_{2j}^{MK}, \quad (8)$$

where $m_{2j}^{MK}$ are the spectral moments of the McKay law, defined in (2). Furthermore, the limiting spectral density of $\mathbf{W}_c$ is given by

$$f_{\mathcal{L}}(x) = \frac{1}{4\pi x^2} \frac{d\sqrt{4(d-1)-\left(d-\frac{1}{2x}\right)^2}}{d^2-\left(d-\frac{1}{2x}\right)^2}, \quad (9)$$

for $x \in \left[\frac{1}{2d+4\sqrt{d-1}}, \frac{1}{2d-4\sqrt{d-1}}\right]$, and $f_{\mathcal{L}}(x) = 0$, otherwise.

## IV. EXAMPLES

To illustrate the applicability of our results we consider two special cases where the state matrices are given by well-known random matrix and graph models.

**Example 1.** *Gaussian-weighted Erdős-Rényi Ensemble*

To illustrate our results, we consider a state matrix with the sparsity structure of an Erdős-Rényi random graph with edge probability $p_n$. Additionally, we assume that the weights of the edges in the random graph are also random. In particular, we assume the edge weights are Gaussian variables with zero mean and variance $\sigma_n^2 = \frac{\alpha^2}{4p_n n}$ for any $0 < \alpha < 1$.[3] We call this random matrix ensemble the Gaussian-weighted Erdős-Rényi Ensemble (GER) with parameters $(n, p_n, \sigma_n^2)$. Let $\mathbf{A}_n$ be an instance from this ensemble. The variance of the entries of $\mathbf{A}_n$ are $p_n \sigma_n^2 = \alpha^2/4n$, and they have zero mean. According to the semicircle law, as $n \to \infty$ the eigenvalue distribution is supported in the range $[-\alpha, \alpha]$; thus, $\mathbf{A}_n$ is Schur stable. The limiting spectral density of GER $(n, p_n, \sigma_n^2)$ can be derived from (1) as $f_{GER}(x) = (2/\pi\alpha^2)\sqrt{\alpha^2-x^2}$, for $x \in [-\alpha, \alpha]$ and 0 otherwise. Fig. 1 shows the histogram of eigenvalues of $\mathbf{A}_n$ from a Gaussian-weighted Erdős-Rényi Ensemble with parameters $n = 1000$, $p_n = 2\log n/n = 0.006$, and $\alpha = 0.5$.

Similarly, the limiting spectral moments of GER are $m_{2k}^{GER} = \left(\frac{\alpha}{2}\right)^{2k} \frac{1}{k+1}\binom{2k}{k}$, and $m_j^{GER} = 0$ for $j$ odd. The trace of the Gramian inverse, $\text{tr}(\mathbf{W}_d^{-1})$, is a Gramian-based control energy metric that quantifies the average steering energy along all directions in the state space. Accordingly, for the controllability Gramian of a random DT system with a state matrix drawn from the GER $(n, p_n, \sigma_n^2)$, we have that asymptotically $\mathbb{E}\{\text{tr}(\mathbf{W}_d^{-1})\} \sim \lim_{n\to\infty} \mathbb{E}\{\text{tr}(I_n - \mathbf{A}_n^2)\}$ $= \lim_{n\to\infty}(n - n\mathbb{E}\{m_2(A_n)\})$. For the particular case of GER, we have that $\mathbb{E}\{m_2(\mathbf{A}_n)\} \sim m_2^{GER} = \alpha^2/4$ so that $\mathbb{E}\{\text{tr}(\mathbf{W}_d^{-1})\} \sim n(1 - \alpha^2/4)$, and the average control energy increases linearly with the increasing size $n$. Notice that the control energy decreases as $\alpha$, which is related to the variance of the edge weights, increases. Fig. 2 shows the histogram of eigenvalues of controllability Gramian for $\mathbf{A}_n$ using the same parameters as in Fig. 1.

---
[3]In fact, we can use any distribution (not necessarily the normal density) that has zero mean, with variance $\sigma_n^2$.

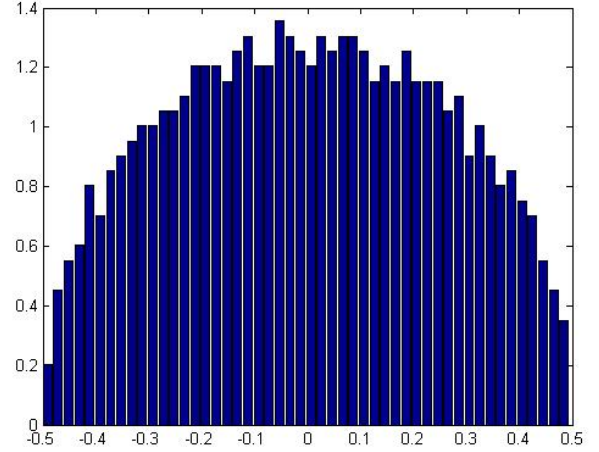

Fig. 1. Eigenvalue Histogram of a Gaussian-weighted Erdős-Rényi ensemble with parameters $n = 1000$, $p_n = 2\log n/\sqrt{n}$, and $\alpha = 0.5$ is given by a semi-circular density that is supported over $[-0.5, 0.5]$.

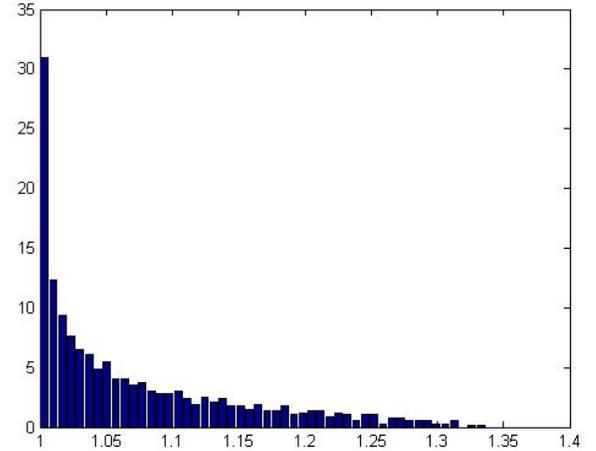

Fig. 2. Eigenvalue Histogram of the Gramian of a discrete-time system with the state-matrix $\mathbf{A}_n$ given by a Gaussian-weighted Erdős-Rényi ensemble with the same parameters as in Fig. 1. The Gramian spectrum in this case follows the limiting spectral density $f_W$ given in Theorem 1.

**Example 2.** *Random Regular Graph Laplacian Dynamics*

As a second example, consider the stabilized Laplacian dynamics on a random $d$-regular graph considered in Theorem 2. In Fig. 3 we have plotted the histogram of the eigenvalues of the stabilized Laplacian $\tilde{\mathbf{L}}_n$ with $d = 3$ and $n = 1000$. The eigenvalue spectrum of the associated controllability Gramian is plotted in Fig. 4. Here, we use $1/\lambda_{\min}(\mathbf{W}_c)$ as a measure of the worst-case minimum required control energy for making a unit transfer in the state space. Applying Theorem 2 we get that $1/\lambda_{\min}(\mathbf{W}_c) \sim 2d + 4\sqrt{d-1}$; hence, *the minimum required energy increases with the increasing degree*.

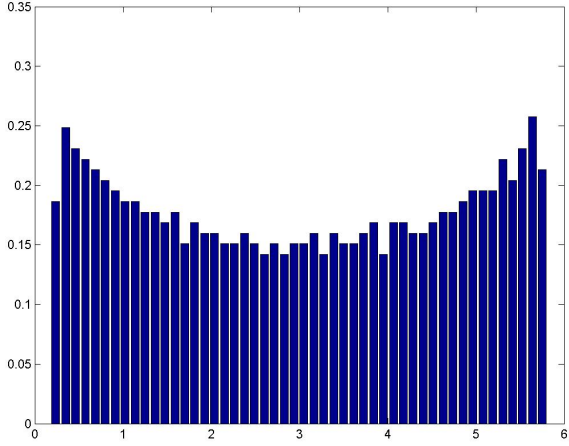

Fig. 3. Eigenvalue Histogram of the stabilized Laplacian $\widetilde{\mathbf{L}}_n := \mathbf{L}_n + \frac{d}{n}\mathbf{1}\mathbf{1}^T$ with $d=3$ and $n=1000$. The limiting spectral distribution in this case is given by the McKay law (Subsection II-C), shifted so that the eigenvalue spectrum is centered around $d=3$.

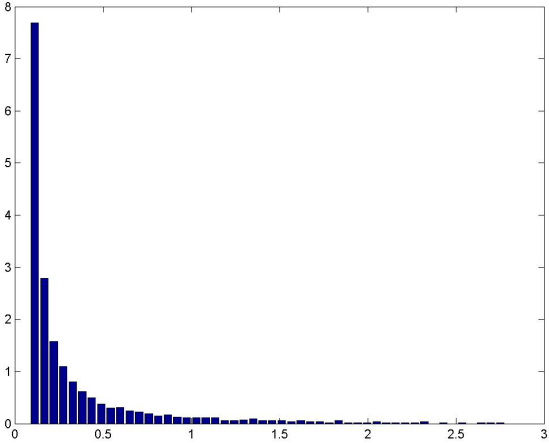

Fig. 4. Eigenvalue Histogram of the Gramian of the stabilized Laplacian dynamics on a random $d$-regular graph with $d=3$ and $n=1000$. The limiting spectral distribution of the Gramian in this case has the density function $f_{\mathcal{L}}$ given in Theorem 2.

## V. CONCLUSIONS

In this paper, we have introduced a theoretical framework to study the eigenvalue spectra of the controllability Gramian of systems with random state matrices. Using tools from random matrix theory, such as the method of moments, we have derived expressions for the spectral moments of the controllability Gramian for both continuous- and discrete-time systems. We have used these moments to derive closed-form expressions for the spectral distribution of the Gramian of random systems derived from popular random matrix and graph ensembles. Finally, we have illustrated how the eigenvalue distribution of the Gramian can be used to draw conclusions about the energy required to control a random system.

## APPENDIX
## PROOFS OF THE MAIN RESULTS

### A. Proof of Lemma 1

Given the input matrix $B_n = I_n$ and a symmetric state matrix $\mathbf{A}_n = \mathbf{A}_n^T$ with spectral moments given by $\mathbf{m}_k := \frac{1}{n}\sum_{i=1}^n \lambda_i(\mathbf{A}_n)^k = \frac{1}{n}\text{tr}(\mathbf{A}_n^k)$, we can write the spectral moments of the DT Gramian $\mathbf{W}_d := \sum_{\tau=0}^\infty \mathbf{A}_n^{2\tau}$, as follows:

$$m_k(\mathbf{W}_d) := \frac{1}{n}\text{tr}(\mathbf{W}_d^k) = \frac{1}{n}\text{tr}\left[\left(\sum_{\tau=0}^\infty \mathbf{A}_n^{2\tau}\right)^k\right]$$
$$= \frac{1}{n}\text{tr}\left[(I - \mathbf{A}_n^2)^{-k}\right].$$

From the Taylor expansion $(I - \mathbf{A}_n^2)^{-k} = \sum_{j=0}^\infty \binom{j+k-1}{k-1}\mathbf{A}_n^{2j}$, we have that:

$$m_k(\mathbf{W}_d) = \frac{1}{n}\text{tr}\left[\sum_{j=0}^\infty \binom{j+k-1}{k-1}\mathbf{A}_n^{2j}\right]$$
$$= \sum_{j=0}^\infty \binom{j+k-1}{k-1}\frac{1}{n}\text{tr}(\mathbf{A}_n^{2j})$$
$$= \sum_{j=0}^\infty \binom{j+k-1}{k-1}\mathbf{m}_{2j},$$

as stated in Lemma 1.

### B. Proof of Theorem 1

We begin by showing that the spectral moments of the Gramian are asymptotically given by (5). Starting form (4) and replacing the moments $\mathbf{m}_k$ of the Wigner ensemble, we obtain

$$\frac{1}{n}\mathbb{E}\left\{\text{tr}[\mathbf{W}_d^k]\right\} = \sum_{j=0}^\infty \binom{j+k-1}{k-1}\sigma^{2j}\frac{1}{j+1}\binom{2j}{j}$$
$$= \sum_{j=0}^\infty \binom{j+k-1}{k-1}(4\sigma^2)^j\frac{2^{-2j}(2j)!}{j!(j+1)!}. \quad (10)$$

Next, we can write $(2j)! = (2^j)(j!)(3 \times 5 \times 7 \times \ldots (2j-1))$ so that

$$\frac{2^{-2j}(2j)!}{j!(j+1)!} = \frac{2^{-j}(3 \times 5 \times 7 \times \ldots (2j-1))}{(j+1)!}$$
$$= \frac{(1/2)(1/2+1)(1/2+2)(1/2+3)\ldots(1/2+j-1)}{2(2+1)(3+1)\ldots(2+j-1)}$$
$$= \frac{(1/2)_j}{(2)_j}, \quad (11)$$

where $(q)_j = q(q+1)\ldots(q+j-1)$. Replacing (11) in (10) yields

$$\frac{1}{n}\mathbb{E}\left\{\text{tr}[\mathbf{W}^k]\right\} = \sum_{j=0}^{\infty}\binom{j+k-1}{k-1}\frac{(1/2)_j}{(2)_j}(4\sigma^2)^j$$

$$= \sum_{j=0}^{\infty}\frac{(k)_j(1/2)_j}{(j!)(2)_j}(4\sigma^2)^j$$

$$= {}_2F_1(\frac{1}{2},k;2,4\sigma^2),$$

when $|4\sigma^2| < 1$, as we wanted to prove.

We now proceed to determine the limiting spectral density of the Gramian when $\mathbf{A}_n$ is a random matrix from the Wigner ensemble. We begin by noting two identities, which hold true for hyper-geometric functions

$${}_2F_1(\frac{1}{2},k;2,4\sigma^2) = (1-4\sigma^2)^{-k}{}_2F_1(k,\frac{3}{2};2,\frac{4\sigma^2}{4\sigma^2-1}), \quad (12)$$

$${}_2F_1(a,b;c;z) = B(b,c-b)^{-1}\int_0^1 \frac{x^{b-1}(1-x)^{c-b-1}}{(1-zx)^a}\,\mathrm{d}x, \quad (13)$$

for $c > b > 0$ and $|z| < 1$, where $B(\cdot,\cdot)$ is the Beta function.[4] The first formula follows from Kummer solutions to Gaussian hyper-geometric differential equation, after a proper change of variables [38, Section 3.7]. The second one is an integral representation of hyper-geometric functions known as Euler formula [38, Section 3.6]. We can use (13) to rewrite the left-hand side of (12) as follows

$${}_2F_1(\frac{1}{2},k;2,4\sigma^2)$$

$$= (1-4\sigma^2)^{-k}\frac{2}{\pi}\int_0^1 x^{\frac{1}{2}}(1-x)^{-\frac{1}{2}}\left(1-\frac{4\sigma^2}{4\sigma^2-1}x\right)^{-k}\mathrm{d}x$$

$$= \frac{2}{\pi}\int_0^1 x^{\frac{1}{2}}(1-x)^{-\frac{1}{2}}\left(1-4\sigma^2+4\sigma^2 x\right)^{-k}\mathrm{d}x$$

$$= -\frac{1}{2\pi\sigma^2}\int_{1/(1-4\sigma^2)}^{1}\sqrt{\frac{4y\sigma^2-y+1}{4y\sigma^2}}\sqrt{\frac{4y\sigma^2}{y-1}}y^k\frac{\mathrm{d}y}{y^2}$$

$$= \int_1^{1/(1-4\sigma^2)}\frac{1}{2\pi\sigma^2 y^2}\sqrt{\frac{4y\sigma^2-y+1}{y-1}}y^k\,\mathrm{d}y$$

$$= \int_{\mathbb{R}}y^k f_W(y)\,\mathrm{d}y,$$

where we have used the fact that $B(\frac{3}{2},\frac{1}{2}) = \pi/2$, together with the change of variables $y^{-1} = 1-4\sigma^2+4\sigma^2 x$, $x = \frac{4y\sigma^2-y+1}{4y\sigma^2}$, and $\mathrm{d}x = -\frac{\mathrm{d}y}{4\sigma^2 y^2}$ for the integration. Hence, the moments of the spectral distribution in (6) are indeed given by the spectral moments in (5). Therefore, the claim in our theorem follows from the method of moments.

---

[4]The Beta function is defined in terms of the Gamma function as $B(x,y) = \frac{\Gamma(c)}{\Gamma(b)\Gamma(c-b)}$.

## C. Proof of Lemma 2

Since the CT Gramian $\mathbf{W}_c$ satisfy (3), its spectral moments are given by

$$m_k(\mathbf{W}_c) = \frac{1}{n}\text{tr}\left[\left(\int_0^{\infty}e^{2\mathbf{A}_n\tau}\mathrm{d}\tau\right)^k\right]$$

$$= \frac{1}{n}\sum_{i=1}^{n}\left(\int_0^{\infty}e^{2\lambda_i(\mathbf{A}_n)\tau}\mathrm{d}\tau\right)^k,$$

Since $\int_0^{\infty}e^{2\lambda_i(\mathbf{A}_n)\tau}\mathrm{d}\tau = -1/(2\lambda_i(\mathbf{A}_n))$, we have that,

$$m_k(\mathbf{W}_c) = \frac{1}{n(-2)^k}\sum_{i=1}^{n}(\lambda_i(\mathbf{A}_n))^{-k}. \quad (14)$$

The Taylor expansion of $x^{-k}$ around $\mathbf{m}_1$ is given by

$$x^{-k} = \sum_{j=0}^{\infty}\binom{-k}{j}(\mathbf{m}_1)^{-k-j}(x-\mathbf{m}_1)^j$$

We can now replace for $(\lambda_i(\mathbf{A}_n))^{-k}$ in (14) to get

$$m_k(\mathbf{W}_c) = \frac{1}{n(-2)^k}\sum_{i=1}^{n}\sum_{j=0}^{\infty}\binom{-k}{j}\frac{(\lambda_i(\mathbf{A}_n)-\mathbf{m}_1)^j}{(\mathbf{m}_1)^{k+j}}$$

$$= \frac{1}{(-2)^k}\sum_{j=0}^{\infty}\binom{-k}{j}\frac{\sum_{i=1}^{n}(\lambda_i(\mathbf{A}_n)-\mathbf{m}_1)^j}{n(\mathbf{m}_1)^{k+j}}$$

$$= \frac{1}{(-2)^k}\sum_{j=0}^{\infty}\binom{-k}{j}\frac{\widetilde{\mathbf{m}}_j}{(\mathbf{m}_1)^{k+j}},$$

as claimed in Lemma 2.

## D. Proof of Theorem 2

Given that $\mathbf{A}_n = -\widetilde{\mathbf{L}}_n$, we can replace for the moments of $\mathbf{A}_n$ in (7). Hence, using the asymptotic moments $\mathbf{m}_1 = \frac{1}{n}\text{Tr}(-\widetilde{\mathbf{L}}_n) \sim -d$ and $\widetilde{\mathbf{m}}_\mathbf{j} = m_j^{MK}$, as discussed in Subsection II-C, (7) becomes

$$\mathbb{E}\{m_k(\mathbf{W}_c)\} \sim \frac{1}{(-2)^k}\sum_{j=0}^{\infty}\binom{-k}{j}(-d)^{-k-j}m_j^{MK}$$

$$= \frac{1}{(2d)^k}\sum_{j=0}^{\infty}\binom{2j+k-1}{2j}\frac{1}{d^{2j}}m_{2j}^{MK}$$

where in the last equality, we have used $\binom{j+k-1}{j}(-1)^j = \binom{-k}{j}$ and substituted $j$ by $2j$ to account for the fact that $m_j^{MK} = 0$ for $j$ odd. This proves the expression claimed in (8) for the asymptotic spectral moments of the CT Gramian.

Once we have a closed-form expression for the expected spectral moments of the Gramian, we need to prove that these moments correspond to those of the density in (9). We begin

by calculating the moments of (9) as follows

$$\int_{\mathbb{R}} x^k f_{\mathcal{L}}(x)\mathrm{d}x$$
$$= \int_{1/(2d+4\sqrt{d-1})}^{1/(2d-4\sqrt{d-1})} \frac{x^k d\sqrt{4(d-1)-(d-\frac{1}{2x})^2}}{4\pi x^2 (d^2-(d-\frac{1}{2x})^2)}\mathrm{d}x$$
$$= \int_{-2\sqrt{d-1}}^{2\sqrt{d-1}} \frac{d\sqrt{4(d-1)-u^2}}{(d-u)^k (2\pi)2^k (d^2-u^2)}\mathrm{d}u, \quad (15)$$

where in the last equality we have used the change of variables $u = d - (1/2x)$. We now use the Taylor expansion of $(d-u)^{-k}$ around $d$, which is given by

$$(d-u)^{-k} = \sum_{j=0}^{\infty} \binom{j+k-1}{j} \frac{u^j}{d^{k+j}}.$$

Using this expansion in (15), we get that

$$\int_{\mathbb{R}} x^k f_{\mathcal{L}}(x)\mathrm{d}x = \frac{1}{2^k} \sum_{j=0}^{\infty} \binom{j+k-1}{j} d^{-k-j} \times \ldots$$
$$\int_{-2\sqrt{d-1}}^{2\sqrt{d-1}} \frac{u^j d\sqrt{4(d-1)-u^2}}{2\pi(d^2-u^2)}du$$
$$\stackrel{\text{a}}{=} \frac{1}{2^k} \sum_{j=0}^{\infty} \binom{2j+k-1}{2j} d^{-k-2j} m_{2j}^{MK}$$
$$= \frac{1}{(2d)^k} \sum_{j=0}^{\infty} \binom{2j+k-1}{2j} d^{-2j} \times \ldots$$
$$\sum_{r=1}^{j} \binom{2j}{r} \frac{2j-2r+1}{2j-r+1} (d-1)^r$$
$$= \mathbb{E}\{m_k(\mathbf{W}_c)\},$$

where in $(\stackrel{\text{a}}{=})$ we have used the fact that $m_j^{MK} = 0$ for $j$. Having thus shown that the moments of $f_{\mathcal{L}}(\cdot)$ coincides with those of the Gramian spectrum the proof follows from the method of moments, described in Section II.

## REFERENCES

[1] R. Kalman, Y. C. Ho, and K. S. Narendra, "Controllability of linear dynamical systems," *Contrib. Differential Equations*, vol. 1, pp. 189–213, 1962.
[2] H. Tanner, "On the controllability of nearest neighbor interconnections," in *Proceedings of the 43rd IEEE Conference on Decision and Control*, Dec. 2004, pp. 2467–2472.
[3] A. Rahmani, M. Ji, M. Mesbahi, and M. Egerstedt, "Controllability of multi-agent systems from a graph-theoretic perspective," *SIAM Journal on Control and Optimization*, vol. 48, no. 1, pp. 162–186, 2009.
[4] S. Martini, M. Egerstedt, and A. Bicchi, "Controllability analysis of multi-agent systems using relaxed equitable partitions," *Int. J. Systems, Control and Communications*, vol. 2, no. 1,2,3, pp. 100–121, 2010.
[5] Z. Ji, Z. Wang, H. Lin, and Z. Wang, "Interconnection topologies for multi-agent coordination under leaderfollower framework," *Automatica*, vol. 45, no. 12, pp. 2857–2863, 2009.
[6] B. Liu, T. Chu, L. Wang, and G. Xie, "Controllability of a leader-follower dynamic network with switching topology," *IEEE Transactions on Automatic Control*, vol. 53, no. 4, pp. 1009–1013, 2008.
[7] G. Parlangeli and G. Notarstefano, "On the reachability and observability of path and cycle graphs," *IEEE Transactions on Automatic Control*, vol. 57, no. 3, pp. 743–748, 2012.
[8] M. Nabi-Abdolyousefi and M. Mesbahi, "On the controllability properties of circulant networks," *IEEE Transactions on Automatic Control*, vol. 58, no. 12, pp. 3179–3184, 2013.
[9] G. Notarstefano and G. Parlangeli, "Controllability and observability of grid graphs via reduction and symmetries," *IEEE Transactions on Automatic Control*, vol. 58, no. 7, pp. 1719–1731, 2013.
[10] S. Zhang, M. K. Camlibel, and M. Cao, "Controllability of diffusively-coupled multi-agent systems with general and distance regular coupling topologies," in *50th IEEE Conference on Decision and Control and European Control Conference (CDC-ECC)*, 2011, pp. 759–764.
[11] A. Chapman, M. Nabi-Abdolyousefi, and M. Mesbahi, "Controllability and observability of network-of-networks via cartesian products," *Automatic Control, IEEE Transactions on*, vol. 59, no. 10, pp. 2668–2679, Oct 2014.
[12] Y.-Y. Liu, J.-J. Slotine, and A.-L. Barabási, "Controllability of complex networks," *Nature*, vol. 473, no. 7346, pp. 167–173, 2011.
[13] M. A. Rahimian and A. G. Aghdam, "Structural controllability of multi-agent networks: Robustness against simultaneous failures," *Automatica*, vol. 49, no. 11, pp. 3149–3157, 2013.
[14] A. Chapman and M. Mesbahi, "On strong structural controllability of networked systems: a constrained matching approach," in *American Control Conference (ACC)*, 2013, pp. 6126–6131.
[15] C. Johnson, "Optimization of a certain quality of complete controllability and observability for linear dynamical systems," *Journal of Fluids Engineering*, vol. 91, no. 2, pp. 228–237, 1969.
[16] P. Muller and H. Weber, "Analysis and optimization of certain qualities of controllability and observability for linear dynamical systems," *Automatica*, vol. 8, no. 3, pp. 237 – 246, 1972.
[17] W. Rugh, *Linear System Theory*. Prentice Hall, 1993.
[18] T. H. Summers, F. L. Cortesi, and J. Lygeros, "On submodularity and controllability in complex dynamical networks," *ArXiv e-prints*, Apr. 2014.
[19] F. Pasqualetti, S. Zampieri, and F. Bullo, "Controllability metrics, limitations and algorithms for complex networks," *Control of Network Systems, IEEE Transactions on*, vol. 1, no. 1, pp. 40–52, March 2014.
[20] V. Tzoumas, M. A. Rahimian, G. J. Pappas, and A. Jadbabaie, "Minimal actuator placement with bounds on control effort," *IEEE Transactions on Control of Network Systems*, 2015, in press.
[21] E. Wigner, "Characteristic vectors of bordered matrices with infinite dimensions," *Ann. of Math.*, vol. 62, pp. 548–564, 1955.
[22] ——, "On the distribution of the roots of certain symmetric matrices," *Annals of Mathematics*, vol. 67, no. 2, pp. pp. 325–327, 1958.
[23] M. Mehta and M. Gaudin, "On the density of eigenvalues of a random matrix," *Nuclear Physics*, vol. 18, pp. 420–427, 1960.
[24] F. J. Dyson, "A brownian-motion model for the eigenvalues of a random matrix," *Journal of Mathematical Physics*, vol. 3, no. 6, pp. 1191–1198, 1962.
[25] G. Anderson, A. Guionnet, and O. Zeitouni, *An Introduction to Random Matrices*, 1st ed. Cambridge: Cambridge University Press, 2009.
[26] T. Tao, *Topics in random matrix theory*. Graduate Studies in Mathematics, American Mathematical Society, 2012.
[27] F. Chung, L. Lu, and V. Vu, "Spectra of random graphs with given expected degrees," *Proceedings of the National Academy of Sciences*, vol. 100, no. 11, pp. 6313–6318, 2003.
[28] F. Chung and M. Radcliffe, "On the spectra of general random graphs," *the Electronic Journal of Combinatorics*, vol. 18, no. 1, p. P215, 2011.
[29] R. R. Nadakuditi and M. E. Newman, "Spectra of random graphs with arbitrary expected degrees," *Physical Review E*, vol. 87, no. 1, p. 012803, 2013.
[30] L. Erdős, A. Knowles, H.-T. Yau, and J. Yin, "Spectral statistics of Erdős–Rényi graphs I: Local semicircle law," *The Annals of Probability*, vol. 41, no. 3B, pp. 2279–2375, 2013.
[31] M. Mariton, "Stochastic controllability of linear systems with markovian jumps," *Automatica*, vol. 23, no. 6, pp. 783–785, 1987.
[32] M. Nabi-Abdolyousefi and M. Mesbahi, "System theory over random consensus networks: controllability and optimality properties," in *IEEE Conference on Decision and Control and European Control Conference (CDC-ECC)*, 2011, pp. 2323–2328.
[33] B. M. Sanandaji, M. B. Wakin, and T. L. Vincent, "Observability with random observations," *IEEE Transactions on Automatic Control*, vol. 59, no. 11, pp. 3002–3007, 2014.

[34] C. Enyioha, M. A. Rahimian, G. J. Pappas, and A. Jadbabaie, "Controllability and fraction of leaders in infinite networks," in *IEEE 53rd Annual Conference on Decision and Control (CDC)*, Dec 2014, pp. 1359–1364.
[35] G. Yan, G. Tsekenis, B. Barzel, J.-J. Slotine, Y.-Y. Liu, and A.-L. Barabasi, "Spectrum of controlling and observing complex networks," *Nature Physics*, vol. 11, no. 9, pp. 779–786, Oct 2014.
[36] T. Tao and V. Vu, "Random matrices: The universality phenomenon for wigner ensembles," *arXiv preprint*, 2012.
[37] B. D. McKay, "The expected eigenvalue distribution of a large regular graph," *Linear Algebra and its Applications*, vol. 40, pp. 203–216, 1981.
[38] Y. L. Luke, *The special functions and their approximations*. Academic press, 1969.